\documentstyle[sprocl]{article}

\def\be{\begin{equation}}
\def\ee{\end{equation}}
\def\bea{\begin{eqnarray}}
\def\eea{\end{eqnarray}}

\begin{document}

\title{NON-FORWARD DOUBLE POMERON EXCHANGE IN QCD}

\author{H. M. NAVELET, R. B. PESCHANSKI}

\address{Service de Physique Theorique, CEA-Saclay,\\
F-91191 Gif-sur-Yvette Cedex, France \\
E-mail: pesch@spht.saclay.cea.fr}

\maketitle\abstracts{ We derive the analytic expression of the two one-loop dipole contributions to the elastic 4-gluon amplitude in QCD for arbitrary transverse momentum. The first one  corresponds to the double QCD pomeron exchange, the other to an order $\alpha^2$ correction to one-pomeron exchange.}

\section{Aims}

It is well-known that the bare pomeron singularity in QCD   \cite {bfkl} is violating the Froissart bound. The computation in the QCD framework of unitarity corrections to the bare pomeron is thus required. The first order correction implies the computation of the two-pomeron contribution to the elastic gluon-gluon amplitude. The aim of our paper is to give the first complete derivation of the analytical expression of the one-loop dipole contribution to the elastic amplitude  in the QCD-dipole picture  \cite {mu2} of BFKL dynamics. The solution for the forward amplitude has been already derived in ref. \cite {nav}, and the non-forward expression conjectured on the basis of assuming conformal invariance. Here we give the full proof of the result and thus of the conformal invariant property. Note that the QCD dipole formulation is known \cite {equiv} to be equivalent at tree level to the  derivation of the BFKL amplitude in terms of Feynman graphs. The idea of our derivation 
is to use this formalism as an effective theory defining the propagation and interaction vertices of two QCD Pomerons, which are colorless compound states of reggeized
gluons in the BFKL representation \cite {bfkl}.

We first introduce the $SL(2,{\cal C})$-invariant formalism for the 4-gluon 
elastic amplitude ${\cal A}_Q (k,k';Y)$ in the BFKL derivation. The solution of the BFKL equation is more easily expressed \cite {lip} in terms of the Fourier  transformed amplitude $f_Q (\rho,\rho';Y)$ given by the relation
\begin {equation}
{\cal A}_Q (k,k';Y)=\frac 1{(2\pi)^5}\int d^2\!\rho d^2\!\rho'\ e^ {i\rho\left(k-\frac Q2 \right)-i\rho'\left(k'-\frac Q2 \right)}
f_Q (\rho,\rho';Y).
\label{2}
\end{equation}
Using the $SL(2,{\cal C})$-invariant formalism, the solution of the BFKL equation reads
\begin {equation}
f_Q (\rho,\rho';Y)=\alpha ^2\ \frac {\mid\rho\rho'\mid}{16} \int dh \ \bar E^h_Q (\rho') E^h_Q (\rho)\ {d(h)}e^{\omega(h)Y},
\label{4}
\end{equation}
where the factor $\alpha ^2$ comes from the coupling to incident dipoles. 

In equation (\ref {4}), the symbolic notation
$
\int dh \equiv \sum^{\infty}_{n=-\infty} \ \int d\nu
$ corresponds to the integration over the $SL(2,{\cal C})$ quantum numbers with $h=i\nu + \frac {1-n}2.$ $E^h_Q (\rho)$ and $\omega(h)$ are, respectively, the 
$SL(2,{\cal C})$ Eigenfunctions and Eigenvalues of the BFKL kernel \cite {lip}. The Eigenvalues read
\begin {equation}
\omega(h) = \frac {\bar \alpha N_c}{\pi} \ \chi (h) \equiv 
\frac {\bar \alpha N_c}{\pi}\ 2\left\{\Psi(1)-\Re \left(\Psi\left(\frac {1+\mid n\mid}2 +i\nu\right)\right)\right\},
\label{6}
\end{equation}
where $\Psi \equiv  (\log\Gamma)^{\prime}.$ The $SL(2,{\cal C})$ Eigenvectors are defined by
\begin {equation}
 E^h_Q (\rho) = \frac {2\pi^2}{\mid \rho \mid b(h)}\int d^2b\ e^{iQ\dot b}\ 
 E^h \left(b-\frac{\rho}2,b+\frac{\rho}2\right), 
\label{8}
\end{equation}
with
\begin {equation}
 E^h \left(b-\frac{\rho}2,b+\frac {\rho}2 \right)=(-)^{h-\tilde h}
\left(\frac{\rho}{b^2-\frac{\rho ^2}4}\right)^h\  \left(\frac{\bar \rho}{\bar b^2-\frac{\bar \rho ^2}4}\right)^{\tilde h}, 
\label{10}
\end{equation}
where $\tilde h = 1-\bar h,$  $b$ is the 2-d impact-parameter, and 
\begin {equation}
d(h) = \left\{\left[\nu ^2 + \frac {(n-1)^2}4\right] \left[\nu ^2 + \frac {(n+1)^2}4\right]\right\}^{-1},\  b(h)=\frac {\pi^3 4^{h+\tilde h -1}}{\frac 12 -h}\ \frac {\gamma(1-h)}{\gamma(\frac 12 -h)},
\label{12}
\end{equation}
where, by definition,  $$\gamma (z) \equiv \frac {\Gamma (z)}{\Gamma (1-\tilde z)}.$$ Note that 
an analytic expression of  the Eigenvectors $E^h_Q (\rho)$ in the mixed representation has been provided \cite {equiv} in terms of a combination of products of two Bessel functions. For simplicity, we did not include the impact factors \cite {lip,vertex}. Note also that the leading contribution to the amplitude (\ref {4}) is the $n=0$ component which corresponds to the BFKL Pomeron.

\section{Formulation}
 The formulation of the general one-loop amplitude in the QCD dipole model
can be written:
\begin{eqnarray}
& &f^{(one-loop)} \left( \rho _{0}\rho _{1};\rho' _{0}\rho' _{1} | Y=y+y'\right)= \frac 1{2!(2\pi)^8}\ \times\nonumber \\
& & \int_0^y d \bar y \int_0^{y'} d \bar y' \int \frac{d^{2}\rho _{a_0} d^{2}\rho _{a_1}d^{2}\rho _{b_0}d^{2}\rho _{b_1}}{\left| \rho
_{a}\ \rho _{b}\right| ^{2}} \frac{d^{2}\rho _{a'_0}d^{2}\rho _{a'_1}d^{2}\rho _{b'_0}d^{2}\rho _{b'_1}}{\left| \rho
_{a'}\ \rho _{b'}\right| ^{2}}\nonumber \\ 
&\times&n_{2}\left( \rho _{0}\rho _{1};\rho _{a_0}\rho
_{a_1},\rho _{b_0}\rho _{b_1} | y-\bar y,\bar y\right) 
\bar n_{2}\left( \rho' _{0}\rho' _{1};\rho' _{a_0}\rho'
_{a_1},\rho' _{b_0}\rho' _{b_1} | y'-\bar y',\bar y'\right)\nonumber \\
& & \ \ \ \ \ \ \ \ \ \ \times \  T(\rho _{a_0}\rho
_{a_1},\rho _{a'_0}\rho
_{a'_1})\ \bar T(\rho _{b_0}\rho _{b_1},\rho _{b'_0}\rho _{b'_1}),
\label{18}
\end{eqnarray}
where $\rho _{0}\rho _{1}$   are the transverse 
coordinates of one of the initially colliding dipoles (resp. $\rho' _{0}\rho' _{1}$ for the second one), $\rho _{a_0}\rho _{a_1}$ and $\rho _{b_0}\rho _{b_1},$ the two interacting dipoles emerging from
the dipole $\rho _{0}\rho _{1}$ after evolution in rapidity (resp. $\rho_{i} \to \rho_{i'}$, for the second one). It is important to notice that one has to introduce the probability distributions $n_{2}(\cdot\cdot\cdot| y-\bar y,\bar y)$ of producing two dipoles after a {\it mixed} rapidity evolution, namely with a rapidity $y-\bar y$ with one-Pomeron type of evolution and  a rapidity $\bar y$ with two-Pomeron type of evolution and then one has to integrate over $\bar y.$ The interaction amplitudes
$T(\rho _{a_0}\rho
_{a_1},\rho _{a'_0}\rho
_{a'_1})$ and $T(\rho _{b_0}\rho _{b_1},\rho _{b'_0}\rho _{b'_1})$ are the elementary two-gluon exchange amplitudes between two colorless dipoles, namely
\begin{equation}
T(\rho _{a_0}\rho_{a_1},\rho _{a'_0}\rho_{a'_1})=  \int d^2q" \ 
e^{i\frac {q"}2 \left(\rho _{a_0}\!+\!\rho_{a_1}\!-\!\rho _{a'_0}\!-\!\rho_{a'_1}\right)}
\ f_{q"} (\rho _{a_0}\!-\!\rho_{a_1},\rho _{a'_0}\!-\!\rho_{a'_1};Y\!=\!0).
\label{19}
\end{equation}
$n^{(2)}$ results from the solution of an evolution equation \cite {mu3,n2}. The solution is a mere extension to the mixed evolution of the one formulated in ref. \cite{pe}, namely:
\begin{eqnarray}
&&n_{2}\left( \left. \rho _{0}\rho _{1};\rho _{a_0}\rho _{a_1},\rho _{b_0}\rho
_{b_1}\right|y-\bar y,\bar y\right) =\frac {\bar \alpha N_C}{\pi}\ \int \frac {dh dh_a dh_b}{\mid \rho_{}\rho_a\rho_b\mid^2}\ \times \nonumber\\
 &&\int d\omega_1 \frac {e^{\omega_1 (y-\bar y)}}
{\omega_1-\omega \left( h\right)}
\int d\omega \frac {e^{\omega y}}
{\omega \left(
h_{a}\right) +\omega \left( h_{b}\right) -\omega }
n_{2}^{h,h_a,h_b} \left(\rho _{0}\rho _{1};\rho _{a_0}\rho _{a_1},\rho _{b_0}\rho
_{b_1}\right),
\label{20}
\end{eqnarray}
 where
\begin{eqnarray}
n_{2}^{h,h_a,h_b} \left(\rho _{0}\rho _{1};\rho _{a_0}\rho _{a_1},\rho _{b_0}\rho
_{b_1}\right)&=& \frac 1{a(h)a(h_{a})a(h_{b})}\int d^{2}\rho _\alpha d^{2}\rho _\beta \ d^{2}\rho _\gamma \nonumber\\
{E}^{h_{a}}{\left( \rho
_{a_0 \alpha} ,\rho _{a_1\alpha} \right) }\ {E}^{h_{b}}{\left(
\rho _{b_0\beta} ,\rho _{b_1\beta} \right) }\!\!&&\!\! {E}^{h}{\left( \rho
_{0 \gamma} ,\rho _{1\gamma} \right) } \ {\bar {\cal R}}^{h,h_a,h_b}_{\alpha,\beta,\gamma},
\label{21}
\end{eqnarray}
with
\begin{equation}
{\cal R}^{h,h_a,h_b}_{\alpha,\beta,\gamma}\equiv
\int \frac {d^{2}r_{0}d^{2}r_{1}d^{2}r_{2}}{\left| r_{01}\ r_{02}\ r_{12}\right| ^{2}}
\ E^{h}{\left( r_{0\gamma},r_{1\gamma}\right)}E^{h_{a}}{\left( r_{0\alpha },r_{2\alpha }\right)
}E^{h_{b}}{\left( r_{1\beta} ,r_{2\beta} \right)
}
\label{22}
\end{equation}
where $\rho =\rho _{0}\!-\!\rho _{1},\rho _{a}=\rho _{a_0}\!-\!\rho _{a_1},\ \rho _{b}=\rho _{b_0}\!-\!\rho _{b_1}.$

Conformal invariance implies
\begin{eqnarray}
{\cal R}^{h,h_a,h_b}_{\alpha,\beta,\gamma}\equiv
\left[\rho_{\alpha\beta}\right]^{h-h_a-h_b}
\left[\rho_{\beta\gamma}\right]^{h_a-h_b-h}
\left[\rho_{\gamma\alpha}\right]^{h_b-h_a-h}\nonumber \\
\left[\bar\rho_{\alpha\beta}\right]^{\tilde h-\tilde h_a-\tilde h_b}
\left[\bar \rho_{\beta\gamma}\right]^{\tilde h_a-\tilde h_b-\tilde h}
\left[\bar\rho_{\gamma\alpha}\right]^{\tilde h_b-\tilde h_a-\tilde h}\ 
g_{3{\cal P}}\left(h,h_a,h_b\right),
\label{24}
\end{eqnarray}
where $g_{3{\cal P}}\left(h,h_a,h_b\right)$ is the celebrated triple Pomeron
coupling as obtained in the QCD dipole model,
namely:
\begin{eqnarray}
g_{3{\cal P}}=
\int \frac {d^{2}r_{0}d^{2}r_{1}d^{2}r_{2}}{\left| r_{01}\ r_{02}\ r_{12}\right| ^{2}}
\left[r_{01}\right]^{h}
\left[\frac {r_{02}}{r_0r_2}\right]^{h_a}\nonumber\\ \times \ 
\left[\frac {r_{12}}{\left(1-r_1\right)\left(1-r_2\right)}\right]^{h_b}\left[\bar r_{01}\right]^{\tilde h}
\left[\frac {\bar r_{02}}{\bar r_0\bar r_2}\right]^{\tilde h_a}
\left[\frac {\bar r_{12}}{\left(1-\bar r_1\right)
\left(1-\bar r_2\right)}\right]^{\tilde h_b}.
\label{26}
\end{eqnarray}

Considering the Fourier transforms (\ref {8}) of the $SL(2,{\cal C})$ Eigenvectors, one writes
\begin{eqnarray}
n_{2}^{h,h_a,h_b} \left(\rho _{0}\rho _{1};\rho _{a_0}\rho _{a_1},\rho _{b_0}\rho
_{b_1}\right)= g_{3{\cal P}}\int d^2q_a d^2q_b d^2Q\ e^{-i\left(
q_{a}b_a + q_{b}b_b + Qb\right)}\ \times
\nonumber\\
 \times E^h_Q (\rho)  E^{h_a}_{q_a} (\rho_a) E^{h_b}_{q_b} (\rho_b)\ 
\delta^{(2)}(Q+q_a+q_b)
\int d^2v d^2w \ e^{-i\left(
(q_a-q_b)\frac v2 + Q\frac w2\right)}\nonumber\\
\left\{\left[v\right]^{-1-h-h_a-h_b}\left[\frac{w-v}2\right]^{-1+h-h_a+h_b}\left[\frac{w+v}2\right]^{-1+h+h_a-h_b}\right\}\times\left\{a.h.\right\},
\label{27}
\end{eqnarray}
where $\rho_{\alpha\beta}=v,$ $\rho_{\alpha\gamma}+\rho_{\beta\gamma} =w,$ $2b_a= \rho_{a0}+ \rho_{a1}, $  $2b_b= \rho_{b0}+ \rho_{1} $ and $b$ is
the overall impact parameter. The notation $\left\{a.h.\right\}$ indicates the {\it anti-holomorphic} part of the bracketed term in the integrand for which the integration variables are complex conjugates and the exponents are replaced by their tilde.

Equivalently, the distribution $n_2$ for the lower vertex is given by the same equation (\ref{27}) by using prime indices.

\section{Calculation}
 Inserting these equations and the definition (\ref{19}) in Eqn.(\ref{18}), the integration over intermediate states and variables yields a drastic simplification due to the appearance of 
quite a few $\delta$-functions. First, integrating over impact parameters,
one gets:
$$
\delta(q_a-q"_a)\delta(q_b-q"_b)\delta(q'_a-q"_a)\delta(q'_b-q"_b).
$$
Then, integrating over the intermediate dipole sizes, one finds
$$
\delta^{(2)}(h_a,h"_a)\delta^{(2)}(h_b,h"_b)\delta^{(2)}(h'_a,1-h"_a)\delta^{(2)}(h'_b,1-h"_b),
$$
where 
 $
\delta^{(2)}(h,h') \equiv \delta_{nn'}\delta(\nu-\nu').$ The integration over $q_a-q_b$ finally gives $\delta(v-v').$

Plugging in the general formula (\ref{18}) the results obtained in 
formulae (\ref{19}) to (\ref{27}) and integrating over $\delta$-fuctions, one gets:
\begin{eqnarray}
f_{Q} (\rho,\rho^{\prime};Y)&=& \alpha ^4\ \left(\frac {\alpha N_C}{%
\pi}\right)^2 \int \! dh dh^{\prime}dh_a dh_b \ g_{3{\cal P}}\left(h,h_a,h_b\right)g_{3{\cal P}}%
^*(h^{\prime},h_a,h_b)  \nonumber \\
&\times&\left[E^h_Q(\rho)\ 
\bar E^{h'}_Q(\rho')\right]\ {\cal H}_Q(h,h') \nonumber \\
&\times&\int_0^y\! d\bar y \int_0^{Y-y}\! d\bar y^{\prime}\int\! d\omega d\omega_1
d\omega^{\prime}d\omega^{\prime}_1 \frac {e^{\omega \bar y + \omega_1 (y-%
\bar y)}} {(\omega(h_a)\!+\!\omega(h_b)\!-\!\omega)(\omega_1\!-\!\omega(h))}
\nonumber \\
\ \ \ \ \ \ &\times&\ \frac {e^{\omega^{\prime}\bar y^{\prime}+
\omega^{\prime}_1 (y^{\prime}-\bar y^{\prime})}}{(\omega(h_a)\!+\!%
\omega(h_b)\!-\!\omega^{\prime})(\omega^{\prime}_1\!-\!\omega(h^{\prime}))}\
, \label{28}
\end{eqnarray}
where, after the change of variables $w=v(1-2t)$ $w'=v(1-2t'),$ 
\begin{eqnarray}
{\cal H}_Q(h,h')=\int d^2vd^2td^2t'\ e^{iQv(t-t')}\ \times\ \ \ \ \ \ \nonumber\\
\left\{v^{h\!-\!h'\!-\!1}t^{-\!1\!+\!h\!+\!h_b\!-\!h_a} (1-t)^{-\!1\!+\!h\!-\!h_b\!+\!h_a}t'^{h_a\!-\!h_b\!-\!h'}(1-t')^{-\!h_a\!+\!h_b\!-\!h'}\right\}\times \{a.h.\}.
\label{30}
\end{eqnarray}
After integration over $v$ one obtains \cite{geronimo}
\begin{eqnarray}
{\cal H}_Q(h,h')= \left(\frac 2{\bar Q}\right)^{h-h'}\left(\frac 2{Q}\right)^{\tilde h-\tilde h'}\ e^{i\frac {\pi}2(h-\tilde h)}\ \gamma(h-h')\int d^2td^2t'\ \times \ \ \ \ \ \nonumber\\
\left\{t^{-1+h+h_b-h_a}(1-t)^{-1+h-h_b+h_a}t'^{h_a-h_b-h'}(1-t')^{-h_a+h_b-h'}\right\}\times\{a.h.\}.
\label{31}
\end{eqnarray}
The remaining integral is of a type which has already been met \cite {dotsenko}. Following the method of paper \cite {geronimo} the result can be  expressed as follows
\begin{eqnarray}
{\cal H}_Q(h,h')&=& \frac {1+\tilde h'-\tilde h}{h-h'} \ \frac {\tilde h'+\tilde h}{1-h-h'} \ \frac {e^{i\frac {\pi}2(h-\tilde h)}}{\sin\pi(2-2h)} \nonumber\\
&\times&\gamma(h-h_a+h_b)\ \gamma(1-h'+h_a-h_b).
\label{32}
\end{eqnarray}

The integral over $h'$ can now easily be performed. The remaining poles at $h'=h$ and $h'=1-h$ give twice the same contribution due to the over completeness relation of the $E_Q^h$ generators \cite {lip}. One finally obtains:

\begin{eqnarray}
f_{Q} (\rho,\rho^{\prime};Y)\!\!\!&=&  \alpha ^4\ \left(\frac {\alpha N_C}{%
\pi}\right)^2\int \! dh dh_a dh_b \ \left\vert \frac {g_{3{\cal P}}
(h,h_a,h_b)} {b(h_a)b(h_b)b(h)} \right\vert^2  \left[E^h_Q(\rho)\ 
\bar E^{h}_Q(\rho')\right]\nonumber \\
&\times&\int_0^y \!d\bar y \int_0^{Y\!-\!y}\! d\bar y^{\prime}\int\! d\omega
d\omega_1 d\omega^{\prime}d\omega^{\prime}_1 \frac {e^{\omega \bar y +
\omega_1 (y-\bar y)}} {(\omega(h_a)\!+\!\omega(h_b)\!-\!\omega)(\omega_1\!-%
\!\omega(h))}  \nonumber \\
\ \ \ \ \ \ &\times&\frac {e^{\omega^{\prime}\bar y^{\prime}+
\omega^{\prime}_1 (y^{\prime}-\bar y^{\prime})}}{(\omega(h_a)\!+\!%
\omega(h_b)\!-\!\omega^{\prime})(\omega^{\prime}_1\!-\!\omega(h))}\ .
\label{36}
\end{eqnarray}

This one-loop amplitude preserves the global conformal invariance of the tree-level BFKL 4-gluon amplitude, since the only scale-dependence on $Q$ is present in the conformal Eigenvectors
$\left[E^h_Q(\rho)\ 
\bar E^{h}_Q(\rho')\right].$  It is worthwhile to notice that we recover at $Q=0$ the forward one-loop amplitude which has been computed by a simpler method in paper \cite {nav}. Conformal invariance at one-dipole loop level, which has been assumed in \cite {nav} is thus now fully proven.

\section{Result}

The integration over rapidity variables yields two different contributions
depending on the sign of the quantity $\omega(h_a)\!+\!\omega(h_b)\!-\!%
\omega(h).$ Indeed for $\omega(h_a)\!+\!\omega(h_b)\!<\!\omega(h),$ the
relevant poles are situated at $\omega\!=\!\omega_1\!=\!\omega^{\prime}\!=\!%
\omega^{\prime}_1 =\omega(h).$ leading to expression (\ref{14}), see further on,  which is
associated with the single Pomeron dependence $e^{\omega(h)\ Y}.$ In the
opposite case, namely $\omega(h_a)\!+\!\omega(h_b)\!>\!\omega(h),$ the
relevant poles are situated at $\omega\!=\!\omega_1\!=\!\omega^{\prime}\!=\!%
\omega^{\prime}_1 =\omega(h_a)\!+\!\omega(h_b).$ The resulting amplitude is
given by (\ref{16})which corresponds to the double-Pomeron energy behaviour $e^{(\omega(h_a)\!+\!\omega(h_b))\ Y}.$ Notice that either
expression depends only on the sum $Y=y+y^{\prime},$ as it should from
longitudinal boost invariance.

The final expressions  read: 
\begin{eqnarray}
f^{({\cal P})}_Q (\rho,\rho^{\prime};Y) &=& \bar \alpha ^4 \int dh \ E^h_Q
(\rho^{\prime}) \bar E^h_Q(\rho)\int dh_a dh_b\ \left\vert \frac {g_{3{\cal P%
}} (h,h_a,h_b)} {b(h_a)b(h_b)b(h)} \right\vert^2  \nonumber \\
& &\frac {e^{\omega(h) Y}}{(\chi (h)-\chi (h_a)-\chi (h_b))^2 }\ ,
\label{14}
\end{eqnarray}
\begin{eqnarray}
f^{({\cal P}\otimes{\cal P})}_Q (\rho,\rho^{\prime};Y) &=& \bar \alpha ^4
\int dh \ E^h_Q (\rho^{\prime}) \bar E^h_Q(\rho)\int dh_a dh_b\ \left\vert 
\frac {g_{3{\cal P}} (h,h_a,h_b)} {b(h_a)b(h_b)b(h)} \right\vert^2  \nonumber
\\
&\times&\frac {e^{\left(\omega(h_a)+\omega(h_b)\right) Y}}{(\chi (h)-\chi
(h_a)-\chi (h_b))^2 }\ ,  \label{16}
\end{eqnarray}
where $g_{3{\cal P}}$ is the triple-QCD-Pomeron vertex (e.g. $1\to 2$ dipole vertex
in the $1/N_C$ limit) which has been recently derived \cite {pe} and evaluated \cite{triplep}.

The two contributions correspond respectively to the one dipole loop
correction to the BFKL Pomeron (\ref{14}) and to the two Pomeron
exchange (\ref{16}). Indeed, the energy dependence
of these contributions is fixed by the asymptotic behaviour in $Y$ of
formulae (\ref{14},\ref{16}), which has been shown \cite {nav} to be, respectively,
in the vicinity of the one-Pomeron and two-Pomeron intercepts. Notice that
the forward amplitudes ($Q=0$) are obtained by replacing $E^h_Q(\rho) \to
(\rho)^{\frac 12-h}.$ Formulae (\ref{14},\ref{16}) are the main results of
this paper..

\end{document}